\documentclass[%
 aip,
 amsmath,amssymb,
 reprint,%
]{revtex4-1}

\usepackage{graphicx}
\usepackage{dcolumn}
\usepackage{bm}
\usepackage[switch,columnwise,running]{lineno}
\usepackage[utf8]{inputenc}
\usepackage[T1]{fontenc}
\usepackage{mathptmx}
\usepackage{etoolbox}
\usepackage{xcolor}
\usepackage[colorlinks=true, linkcolor=blue, citecolor=blue, urlcolor=blue]{hyperref}  
\usepackage{flafter}
\makeatletter
\def\@email#1#2{%
 \endgroup
 \patchcmd{\titleblock@produce}
  {\frontmatter@RRAPformat}
  {\frontmatter@RRAPformat{\produce@RRAP{*#1\href{mailto:#2}{#2}}}\frontmatter@RRAPformat}
  {}{}
}%
\makeatother

\begin{document}
 
\preprint{AIP/123-QED}

\title[]{Shearless and periodic attractors in the dissipative Labyrinthic map}
\author{L. F. B. Souza}
 \homepage{luisfernando.bernardi1998@gmail.com}
\affiliation{Institute of Physics, University of São Paulo, São Paulo, 13506-900, SP, Brazil}

\author{R. Egydio de Carvalho}
 \homepage{ricardo.egydio@unesp.br}
\affiliation{%
Department of Statistics, Applied Mathematics and Computer Science, São Paulo State University, Rio Claro, 13506-900, SP, Brazil
}%

\author{R. L. Viana}
 \homepage{rlv640@gmail.com }
\affiliation{%
Universidade Federal do Paraná, Centro Interdisciplinar de Ciência, Tecnologia e Inovação, Núcleo de Modelagem e Computação Científica, Curitiba, 81531-990, PR, Brazil
}%

\author{I. L. Caldas}
 \homepage{ibere@if.usp.br }
\affiliation{%
Institute of Physics, University of São Paulo, São Paulo, 13506-900, SP, Brazil
}%


\begin{abstract}
The Labyrinthic map is a two-dimensional area-preserving map that features a robust transport barrier known as the shearless curve. In this study, we explore a dissipative version of this map,examining how dissipation affects the shearless curve and leads to the emergence of quasi-periodic or chaotic attractors, referred to as shearless attractors. We present a route to chaos of the shearless attractor known as the Curry-Yorke route. To investigate the multi-stability of the system, we employ basin entropy and boundary basin entropy analyses to characterize diverse scenarios. Additionally, we numerically investigate the dynamic periodic structures known as "shrimps" and "Arnold tongues" by varying the parameters of the system.
\end{abstract}
\maketitle
\begin{quotation}
Area-preserving maps are an important means of understanding how systems evolve and how they transport energy or particles. When we look at a Hamiltonian system, a type of system that conserves energy, a perturbation can lead to both periodic and chaotic behavior in its motion simultaneously. An essential propriety in those systems is the twist condition, which helps us distinguish between two types of systems: twist and non-twist. This distinction is key because it influences how a system transitions into chaos. Non-twist systems can reorganize the structure of their phase space, presenting strong barriers that can prevent transport, like shearless curves. These concepts help explain various phenomena, such as wind patterns in the atmosphere and the behavior of magnetic fields in devices that contain plasma. The non-twist maps serve as useful models for studying these dynamics. Additionally, when we introduce dissipation into these systems, we can see the emergence of attractors, which are solutions that can be periodic, quasi-periodic, or chaotic. This study investigates how shearless barriers behave when dissipation is present in a non-twist map, focusing on how their stability is related to the control parameters and the symmetry of the system.
\end{quotation}

\section{Introduction}

Area-preserving maps, a common type of conservative system studied in Hamiltonian dynamics, are widely used to analyze system evolution and transport properties. When considering a slightly perturbed integrable Hamiltonian system, its phase space can exhibit both chaotic and periodic motions, especially under certain parameter conditions. The Poincaré-Birkhoff and KAM theorems provide crucial insights into the behavior of nearly integrable Hamiltonian systems \cite{zaslavsky2005, lichtenberg}.

The non-degeneracy criterion for a Hamiltonian system, described in action and angle coordinates $(I, \theta)$, is expressed as $\partial^{2} H/\partial I^{2} \ne 0$. This criterion, known as the twist condition for maps, is denoted by $\lvert \frac{\partial \theta_{n+1}}{\partial I} \rvert \ne 0$. Therefore, satisfying the twist condition means that the system has non-degenerate orbit frequencies. The transition to chaos in non-twist systems significantly differs from that in twist systems, as the fulfillment of the twist condition is essential for the validity of the KAM theorem. The KAM theorem provides solutions for quasi-integrable Hamiltonian systems with non-degenerate frequency orbits. However, for degenerate Hamiltonian systems, the solution remains an open question.

Moreover, the violation of the twist condition leads to topological reorganizations in phase space. In terms of transport investigation, this violation results in the emergence of a robust transport barrier known as the shearless curve \cite{weele1988, diego1993} or multiple shearless barriers \cite{grime2023BNM1}, along with other related phenomena \cite{corso1998, diego1996, egydio1992, wurm2004}. Various physical phenomena, such as atmospheric zonal flow \cite{delCastilloNegrete2000} and the behavior of magnetic field lines in plasma confinement devices with reversed magnetic shear \cite{Morrison2000, OsorioQuiroga2023, MacKay1984, BernardiDeSouza2022, Oda1995}, can be effectively explained through non-twist systems. Numerical results suggest that the orbit violating the twist condition, the shearless curve, is likely to be the last torus to be destroyed \cite{Wurm2005, Shinohara1998, Szezech2009}.

All non-twist phenomena, except for multiple shearless curves, can be encompassed by the non-twist standard map (SNM) introduced in 1993 \cite{diego1993}. This two-dimensional perturbed symplectic map deviates from the twist condition in a single orbit, featuring only one shearless curve, and is governed by two control parameters. Since its introduction, the SNM has undergone various modifications and expansions \cite{Martins2010, Mugnaine2020, Wurm2013}. One notable modification is the Labyrinthic map (LM) \cite{Martins2010}, another two-dimensional perturbed symplectic map. Unlike the SNM, the LM incorporates two perturbed terms and introduces an additional control parameter that significantly influences the system's dynamics.

Conservative systems inherently maintain the conservation of phase space volume. However, the introduction of dissipation disrupts this conservation, leading to the loss of the previously preserved volume. Consequently, any given initial condition (IC) gradually converges asymptotically to states known as attractors, which can be periodic, quasi-periodic, or chaotic. This is particularly important for modeling practical systems, where dissipation due to friction or drag is always present, making Hamiltonian systems with dissipation more representative of actual physical processes. Previous studies have detailed the impact of dissipation on the LM \cite{deCarvalho2015, Mugnaine2021, Kato2019}, particularly on the shearless torus. These studies revealed the evolution of this torus into an attractor, known as a shearless attractor (SA), which can display quasi-periodic or chaotic behavior depending on the control parameters.

Attractors can undergo abrupt and significant changes in their stability based on the control parameters, one of which is the transition from a quasi-periodic attractor to a chaotic one, known as the Curry-Yorke route \cite{Grebogi1982, Grebogi1983}. These changes are associated with the dynamics of the manifolds and the attractors, such events, often called crises, have been observed in a wide range of experimental and numerical setups \cite{Chizhevsky2000, Ditto1989, dePaula2006, Baroni2021, Nogueira2018}.

In this study, we choose the LM to investigate the evolution of the shearless attractor, with particular emphasis on the map's symmetry, focusing on results in terms of symmetry, as previous evaluations of the shearless attractor were conducted for symmetric non-twist maps with only one perturbative term \cite{Mugnaine2021}. We evaluate the shearless breakup scenarios in the conservative case and, for the same range of parameters, compare the behavior of the shearless curve upon the introduction of dissipation. Subsequently, we present a route to chaos for the shearless curve, where the quasi-periodic attractor transitions into a chaotic band and eventually into a chaotic attractor, following the Curry-Yorke route. We then analyze multi-stability scenarios in the phase space using basin entropy and boundary basin entropy. Finally, we identify and analyze periodic structures in the parameter spaces, such as Arnold tongues and shrimps, which have been observed in several dissipative systems \cite{Gallas1994, Brugnago2011, deOliveira2019, Bonatto2005}. These structures are associated with periodic windows embedded within chaotic regions and are linked to the transition to chaos via two types of bifurcations \cite{Gallas1993, Fraser1982, Baptista1996}, such periodic windows have been observed in Chua’s circuit experiments \cite{Maranhao2008, Gouvea2016}.

This paper is organized as follows: in Sec.\,\ref{section2}, we present the LM main characteristics and discuss the shearless curve breakup in terms of the parameters. In section\,\ref{section3}, we introduce the dissipative version of the LM and abbreviate the map as DLM; we also present the evolution of the orbits when subjects dissipation. In sections\,\ref{section4}, \ref{section5}, \ref{section6}, and \ref{section7}, we discuss the route to chaos, stability, multi-stability, and periodic structures of the system, and the paper is concluded in Sec.\,\ref{section8}.

\section{Basic Characteristics of the LM}\label{section2}

The Labyrinthic map (LM) is an example of a map that captures the dynamic of a nearly integrable Hamiltonian system. The equations define the LM 
\begin{subequations}
\label{Eq1}
    \begin{align}
    \label{Eq1a}
        I_{n+1} &= I_{n} - b\sin{(2\pi \theta_{n})} - b\sin{(2 \pi \eta \theta_{n})},\\
        \theta_{n+1} &= \theta_{n} - a(I_{n+1} - r_1)(I_{n+1} - r_2) \quad \mathrm{mod}\,1,
        \label{Eq1b}
    \end{align}
\end{subequations}
where $I_{n} \in \mathbb{R}$ and $\theta_{n} \in [0,1]$. The control parameters $a$ and $b$ govern the influence of the non-twist term (the function that scales with $a$) and the amplitude of the system’s nonlinearity, respectively. In order to maintain the symplectic propriety of the map, we assume integer values for $\eta$. Furthermore, $r_1$ and $r_2$ represent the positions of primary resonances. For this analysis, we have chosen to set $r_1 = 1$ and $r_2 = -1$. One way to distinguish orbits in phase space is by their frequency, which can be ascertained using the winding number, defined as
\begin{equation}
\label{Eq2}
    \omega = \lim_{n \to \infty}\dfrac{\theta_{n} - \theta_{0}}{n}
\end{equation}
where $n$ the number of iteration, $\theta_{0}$ is the initial condition (IC) and $\theta_{n}$ is the $n$th iteration of $\theta$. Its profile concedes information about the shearless curve since the coordinates of the extrema provide a condition over the shearless torus in the phase space.

In dynamic systems, an important aspect is the presence of symmetry transformations and the fulfillment of a symmetry transformation group. In non-twist systems, when a map meets specific symmetry transformation criteria, it has indicator points, which are conditions on the shearless curve. A transformation $T$ is deemed a symmetry if $M = T^{-1}MT$ holds, indicating that the map remains invariant under $T$. For the map to have indicator points, it must adhere to the following symmetry transformation,

\begin{equation}
S:
    \begin{cases}
        I_{n+1} = -I_n \\
        \theta_{n+1} = \theta_n \pm \frac{1}{2}
    \end{cases}
\end{equation}

the LM is symmetric for the $S$ transformation only for odd values of $\eta$. The indicator points for the LM are given by

\begin{subequations}\label{IP}
\label{Eq3}
    \begin{align}
    \label{Eq 3a}
        (\textbf{\textrm{IP}}_0)_{1,2}&=\left[\pm \dfrac{1}{4}, \dfrac{b}{2}\left(\pm 1 + (-1)^{(\eta-1)/2}\right)\right], \\
          (\textbf{\textrm{IP}}_1)_{3,4}&=\left[-\dfrac{ar_1r_2}{2}\pm \dfrac{1}{4} ,0\right],
          \label{Eq 3b}
    \end{align}
\end{subequations}

\begin{figure*}
     \centering
    \includegraphics[width=\textwidth]{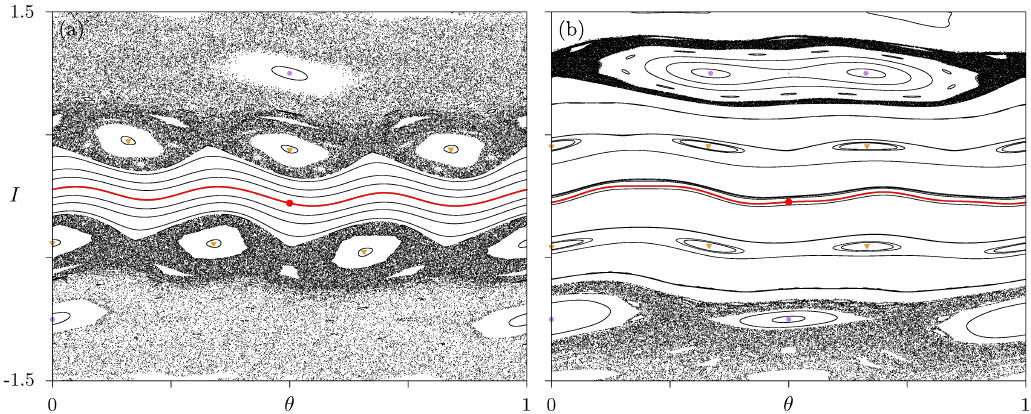}
     \caption{Phase space of the LM for $a = 0.4$ and $b = 0.07$. The red curve is the shearless torus. The red point is the initial condition in (a) $(\theta, I) = (0.5, -0.05576)$ for $\eta=3$ and (b) $(\theta, I) = (0.5, -0.04585)$ for $\eta=2$.}
     \label{fig:1}
\end{figure*}

In Figures \ref{fig:1}\textcolor{blue}{a} and \ref{fig:1}\textcolor{blue}{b}, we illustrate the phase space of the LM for $a=0.4$, $b=0.07$, and $\eta=3$ for Figure \ref{fig:1}a, and $\eta=2$ for Figure \ref{fig:1}\textcolor{blue}{b}. These plots reveal several key features. The resonance islands around the elliptic points (shown as colored points), occur when the frequencies of the perturbation $(b\neq 0)$ become commensurate with the non-perturbative system ($b=0$) frequencies, leading to the formation of regular, closed orbits in phase space. The primary resonance islands at $I=\pm 1$ persist in both cases. However, for $\eta=2$, the primary resonances undergo a saddle-node bifurcation, where the main resonance splits into two distinct resonances. Additionally, we observe chaotic orbits in the regions surrounding these islands. Furthermore, period three islands emerge around $I \approx \pm 0.5$.

Moreover, a series of invariant spanning curves around $I = 0$ delineates stability regions within the system, notably, in red, the shearless curve. The coordinates of a red dot along this curve are specified in Figures \ref{fig:1}\textcolor{blue}{a} and \ref{fig:1}\textcolor{blue}{b}. In Figure \ref{fig:1}\textcolor{blue}{a}, the coordinates are $(\theta, I)=(0.5,-0.05576)$, while in Figure \ref{fig:1}\textcolor{blue}{b}, they are $(\theta, I)=(0.5,-0.04585)$. This red dot represents the initial condition iterated to generate the shearless curve.

\begin{figure*}
     \centering
      \includegraphics[width=\textwidth]{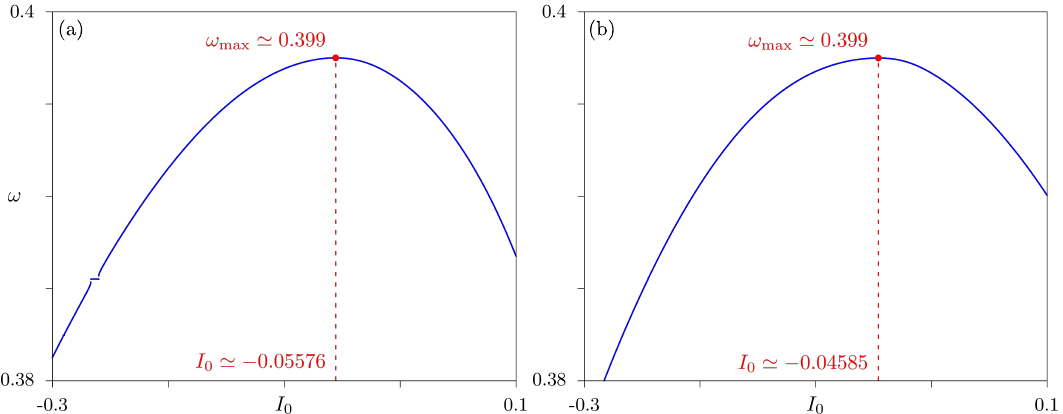}
     \caption{The winding number profile using $\theta_0 = 0.5$ and $y \in [-0.3, 0.1]$. The red dotted line corresponds to the extremum of the winding number. For $a = 0.4$ and $b = 0.07$, in (a) $\eta=3$ and (b) $\eta=2$.}
     \label{fig:2}
\end{figure*}

The initial conditions were determined by locating the maximum point of the winding number profile, as depicted in Figures \ref{fig:2}\textcolor{blue}{a} and \ref{fig:2}\textcolor{blue}{b}, corresponding to $\eta=3$ and $\eta=2$, respectively. Despite variations in $\eta$, the shearless curves exhibit consistent frequencies, $\omega \approx 0.399$. This observation underscores the robustness of the shearless dynamics, persisting across different parameter values.
\begin{figure*}
     \centering
     \includegraphics[width=\textwidth]{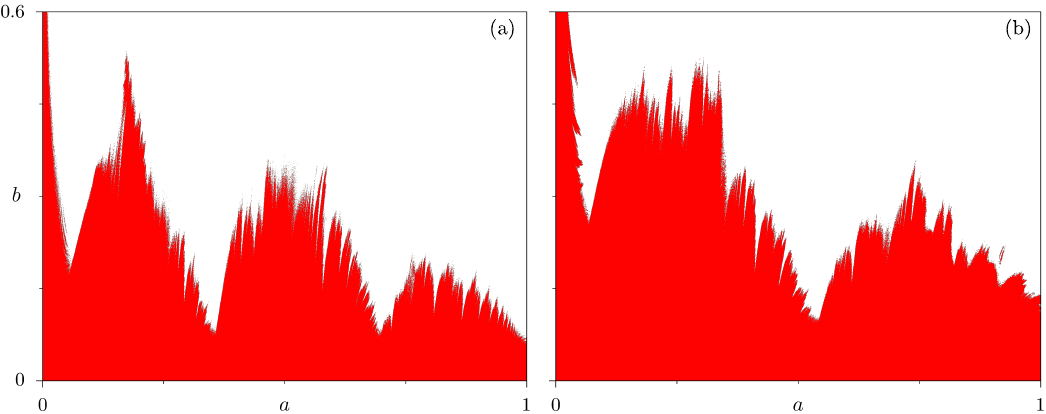}
     \caption{Parameter spaces depicting shearless breakup in (a) for $\eta=3$ based (b) $\eta=2$ In red are scenarios where the shearless curve is present, and in white are scenarios where the shearless curve is destroyed.}
     \label{fig:3}
\end{figure*}

As mentioned, a direct consequence of the choice of $\eta$ is the existence of indicator points. In non-twist systems, the shearless curve is the invariant torus candidate to be the last torus destroyed, therefore studying its breakout in terms of the control parameters of the map gives an insight into the transition to chaos. We choose the crossing flight to identify shearless breakup. Essentially, this involves creating a set of initial conditions in a specific phase space region and iterating each condition a certain number of times. If the orbit of any initial condition reaches a different region of phase space, it indicates that the shearless curve has been destroyed for this set of control parameters.

The parameter spaces $(a,b)$ shown in Figures~\ref{fig:3}\textcolor{blue}{a} and \ref{fig:3}\textcolor{blue}{b} were obtained using the crossing flight criterion for $\eta=3$ and $\eta=2$, respectively. In red, scenarios depict pairs of parameters where shearless exists, while in white, scenarios represent instances where shearless has been destroyed. In this study, $10^3$ initial conditions at $I = 10$ underwent $10^5$ iterations each. The shearless curve was considered destroyed (is no longer an invariant torus, became a chaotic orbit) if the orbit of any initial condition reached $I = -10$. The parameter spaces have similar characteristics, including smooth and non-smooth boundaries. The smooth boundaries are associated with bifurcation phenomena \cite{Wurm2005}. Notably, this criterion is independent of the knowledge of a condition on the shearless curve, making it a recommended method for maps lacking the previously mentioned symmetries. In section~\ref{section6}, we will analyze for the same set of parameters in Fig.~\ref{fig:3}, the dynamics when dissipation is introduced.

\section{Dissipative Labyrinthic Map} \label{section3}

A dissipative version of the LM is defined by
\begin{subequations}
\label{Eq 4}
    \begin{align}
    \label{Eq 4a}
         I_{n+1} &= (1-\gamma)I_{n} - b\sin{(2\pi \theta_{n})} - b\sin{(2 \pi \eta \theta_{n})},\\
         \theta_{n+1} &= \theta_{n} - a(I_{n+1} - r_1)(I_{n+1} - r_2) \quad \mathrm{mod}\,1
        \label{Eq 4b}
    \end{align}
\end{subequations}
where $\gamma$ is the control parameter responsible for the dissipation, for $\gamma = 0$ we recover the conservative case.

When considering dissipation in symplectic maps, the elliptic fixed points (the centers of the islands) become stable focus (sinks), where the trajectories approach asymptotically, forming spirals. The region previously occupied by the islands becomes the basin of attraction (or part of it) of the stable focus, and the chaotic orbits become either chaotic transients that converge towards the focus or (more rarely) orbits belonging to a chaotic attractor. The shearless curve transforms into the shearless attractor (SA), exhibiting either quasi-periodic or chaotic behavior dependent upon control parameters, and it is possible to obtain scenarios with more than one attractor in phase space. The occurrence of multiple attractors is referred to as a multi-stability phenomenon \cite{Feudel1996, Feudel1997, Feudel1998}.  

\begin{figure*}
     \centering
    \includegraphics[width=\textwidth]{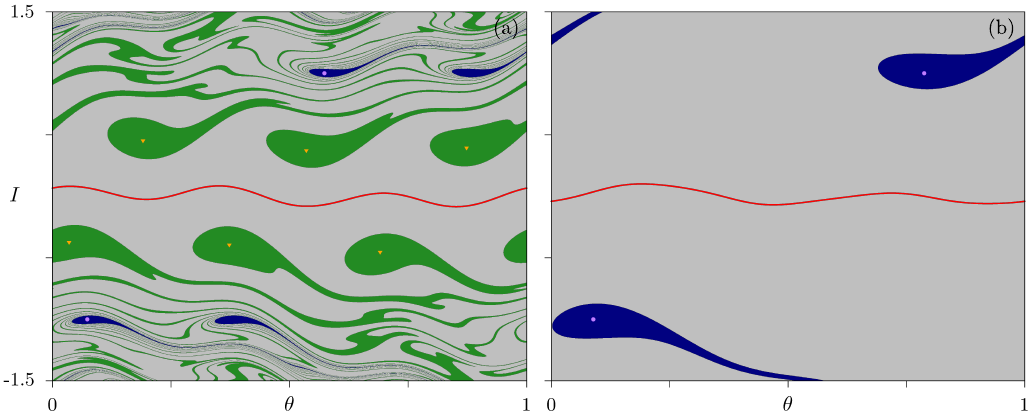}
     \caption{Attractors of the DLM for $a = 0.4$, $b = 0.07$, and $\gamma = 0.1$. In (a) for $\eta=3$ and (b) $\eta=2$. The red curve is the shearless attractor, and the colored dots are periodic attractors.}
     \label{fig:4}
\end{figure*}

Figure \ref{fig:4} illustrates the attractors and their respective basins of attraction when a dissipation parameter $\gamma=0.1$ is introduced for $\eta=3$ and $\eta=2$, corresponding to the same $a$ and $b$ parameter values as in Figure \ref{fig:1}. The shearless curve evolves into a quasi-periodic attractor (the shearless attractor, shown in red). Some of the elliptic points become point attractors (colored dots), and the islands surrounding these elliptic points become their respective basins (represented in various colors).

In Figure \ref{fig:1}a, the phase space shows two primary resonances and a period-three island near the shearless region. When dissipation is introduced, as seen in Figure \ref{fig:4}\textcolor{blue}{a}, the period-three fixed point evolves into a period-three attractor (yellow triangular dots), with the surrounding islands becoming their basins (in green). The same transformation occurs for the period-one attractor. Additionally, the shearless curve evolves into a quasi-periodic attractor (red), with its respective basin shown in gray. The same analysis applies to Figure \ref{fig:4}\textcolor{blue}{b}.

\begin{figure*}
     \centering
    \includegraphics[width=\textwidth]{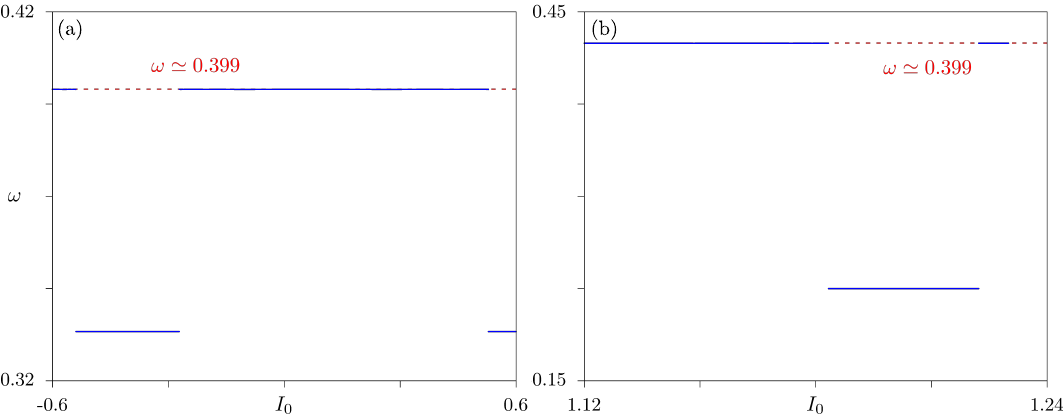}
     \caption{The winding number profile using $\theta_0 = 0.5$. The red dotted line corresponds to the extremum of the winding number. For $a = 0.4$ and $b = 0.07$, in (a) $\eta=3$ and (b) $\eta=2$.}
     \label{fig:5}
\end{figure*}

Figure~\ref{fig:5}, we compute the winding number using Eq.~\ref{Eq2}, disregarding the transient phase (first $1\times 10^{5}$ iterations) to ensure convergence of initial conditions towards an attractor. In both cases, we observe that certain initial conditions converge to an attractor with a winding number close to $\omega \approx 0.399$, consistent with the shearless curve in the conservative scenario. This suggests that the shearless curve transitions into a quasi-periodic attractor upon the introduction of dissipation. These certain initial conditions lie in the shearless attractor basin as expected. Conversely, in the second scenario, initial conditions converge from a different winding number than $\omega \approx 0.399$, indicating convergence towards a periodic attractor. This aligns with the expected evolution of elliptic points in the conservative case towards periodic attractors.

\section{Shearless attractor bifurcation}\label{section4}

\begin{figure*}
    \centering
    \includegraphics[width=0.8\textwidth]{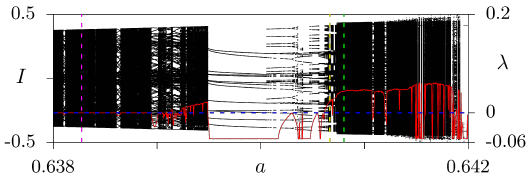}
    \caption{Bifurcation diagram and the Lyapunov exponent vs the parameter a for $b = 0.2$ and $\gamma = 0.1$. The quasi-periodic and periodic attractors alternate between each other until a chaotic behavior appears on the right side of the diagram. The black and red points indicate the diagram and the Lyapunov exponent. The magenta, yellow, and green vertical dashed lines correspond to the values of parameters: $a = 0.638272$, $a = 0.640665$, and  $a = 0.6408$, respectively.}
    \label{fig:6}
\end{figure*}

To analyze the topological changes within the shearless attractor, we compute the bifurcation diagram along with the largest Lyapunov exponent, denoted as $\lambda$, by varying the control parameter $a$ while keeping $(b, \gamma)$ constant. In Fig.\,\ref{fig:6}, we select the initial condition (IC) corresponding to the indicator point \( (\textbf{\textrm{IP}}_0)_{1} \), as defined by Eq.\,\ref{Eq3}. This choice is made assuming that this indicator point likely resides within the basin of attraction of the shearless attractor when dissipative effects are accounted for. 

To compute the Lyapunov exponent, we used the Wolf algorithm that employs the concept of finite-time Lyapunov exponents to estimate the spectrum of Lyapunov exponents \cite{Wolf1985}, capturing the exponential divergence rates of nearby trajectories in chaotic systems through a carefully designed algorithmic approach and is defined as

 \begin{equation}
     \lambda_{j} = \lim_{n \to \infty}\left[\dfrac{1}{n}\ln{\left|\Lambda^{(n)}_{j}\right|}\right], \quad j = 1,2,...,
 \end{equation}
where $\Lambda^{(n)}_{j}$ are the eigenvalues of the matrix $T = \prod^{n}_{i=1} J(\theta_{i},I_{i})$, $J(\theta_{i},I_{i})$ is the Jacobian matrix calculated at the point $(\theta_{i},I_{i})$.  If the orbits diverge, the largest Lyapunov exponent is positive, $\lambda > 0$, and the dynamics are chaotic. If the orbits stay close to each other, the largest Lyapunov exponent is $\lambda = 0$ or $\lambda < 0$, characterizing quasi-periodic and periodic motion, respectively. If, by chance, the exponent profile abruptly goes to zero, the orbit has suffered a bifurcation. 

\begin{figure}
     \centering
      \includegraphics[width=0.45\textwidth]{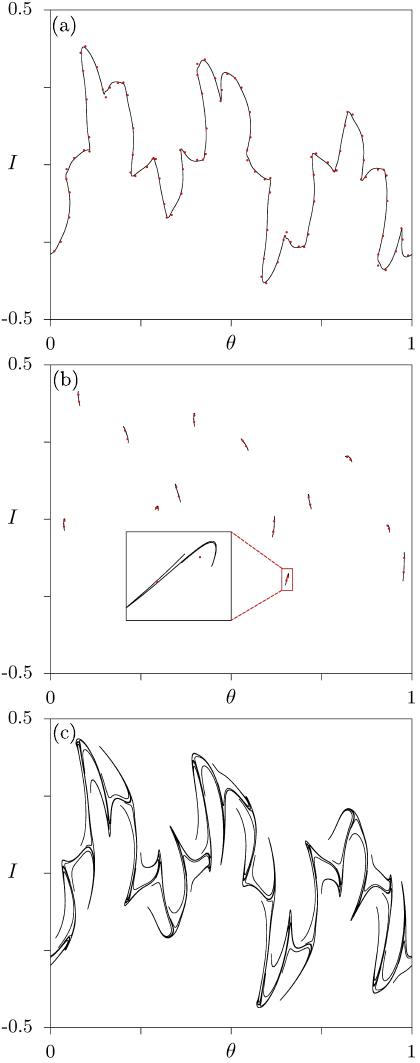}
     \caption{The route to chaos for $b = 0.2$. For (a) $a = 0.638272$, the smooth quasi-periodic attractor on the torus is represented by the black curve, and there is a ghost periodic attractor, indicated by the red points. In (b), we have $a = 0.640665$; there are chaotic banded attractors, and the ghost attractors are two chains of period $40$. For (c) $a = 0.6408$, there is only one chaotic attractor on the torus. The ghosts in (a) and (b) are shown for $a = 0.63862$ and $a = 0.640441$, respectively.}
     \label{fig:7}
\end{figure}

The bifurcation diagram depicted in Fig.\,\ref{fig:6} illustrates a variety of distinct behaviors. Notably, the attractors manifest across a range of values for the variable $I$, with periodic attractors identifiable as discrete points. In the context of Fig.\,\ref{fig:6}, for lower values of parameter $a$, the attractors fill the $I$ range and exhibit a Lyapunov exponent of zero, indicating the presence of quasi-periodic shearless attractors. At the same time, the points represent periodic attractors characterized by a negative Lyapunov exponent. Conversely, as parameter $a$ increases, we observe the emergence of attractors and chaotic windows with a positive Lyapunov exponent. To examine the configurations of these attractors, we focus on three specific points $(a, b = 0.6)$ denoted by vertical dashed lines of purple, yellow, and green colors, constructing corresponding phase spaces for each scenario, as depicted in Figure~\ref{fig:7}.

In Figure~\ref{fig:7}, distinct phenomena are visible when examining the attractors in each phase space. Figure~\ref{fig:7}\textcolor{blue}{a} shows an attractor that resembles a torus, signifying a quasi-periodic shearless attractor surrounded by ghost-fixed points. These ghost points, highlighted in red, correspond to periodic attractors at a different parameter setting, specifically when \( a = 0.7063 \), both displaying a period of $48$. With a slight increase in the parameter  \( a \), Figure~\ref{fig:7}\textcolor{blue}{b} illustrates a different scenario. Multiple separated structures appear in the phase space near the ghost periodic attractors from the previous parameter. The magnified view in Fig.~\ref{fig:7}\textcolor{blue}{b} reveals that each black-separated structure, or band, is traversed by a single orbit, resulting in a phase-locking phenomenon. The positive Lyapunov exponent for case (b) in Fig.\ref{fig:6} (indicated by the yellow vertical dashed line) confirms the presence of a chaotic banded attractor.

Moving to the final phase space, Fig.~\ref{fig:7}\textcolor{blue}{c}, a more intricate attractor shape is observed, indicative of chaotic behavior as confirmed by the Lyapunov exponent in Fig.~\ref{fig:6} (green dashed line). This succession in the phase space illustrates the transition to chaos, wherein an attractor on a "torus" destabilizes, leading to a series of phase-locking (banded attractor) events preceding the emergence of a chaotic attractor.  In Figure~\ref{fig:7}, we witness the Curry-Yorke mechanism, often referred to as the "soft" transition to chaos. This process entails the destabilization of the attractor on a torus, followed by a succession of phase-locking events that lead to a banded attractor before ultimately culminating in chaos; this route to chaos was already reported for the dissipative non-twist map \cite{Mugnaine2021}.
\begin{figure*}
     \centering
    \includegraphics[width=\textwidth]{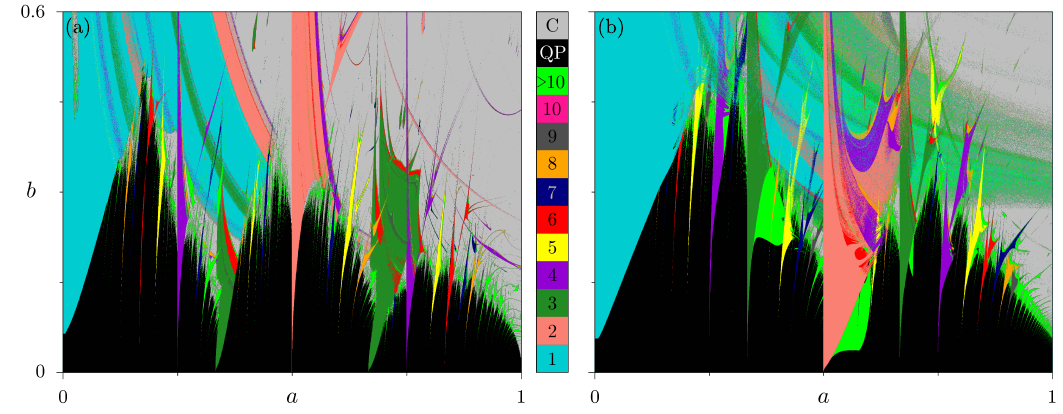}
     \caption{Parameter spaces $(a,b)$ in (a) $\eta=3$ and (b) $\eta=2$. The periods were counted up to $100$ and are represented in color, according to the caption. Chaos (C) and quasi-periodicity (QP) were identified through the largest Lyapunov exponent.}
     \label{fig:8}
\end{figure*}
\begin{figure*}
     \centering
     \includegraphics[width=\textwidth]{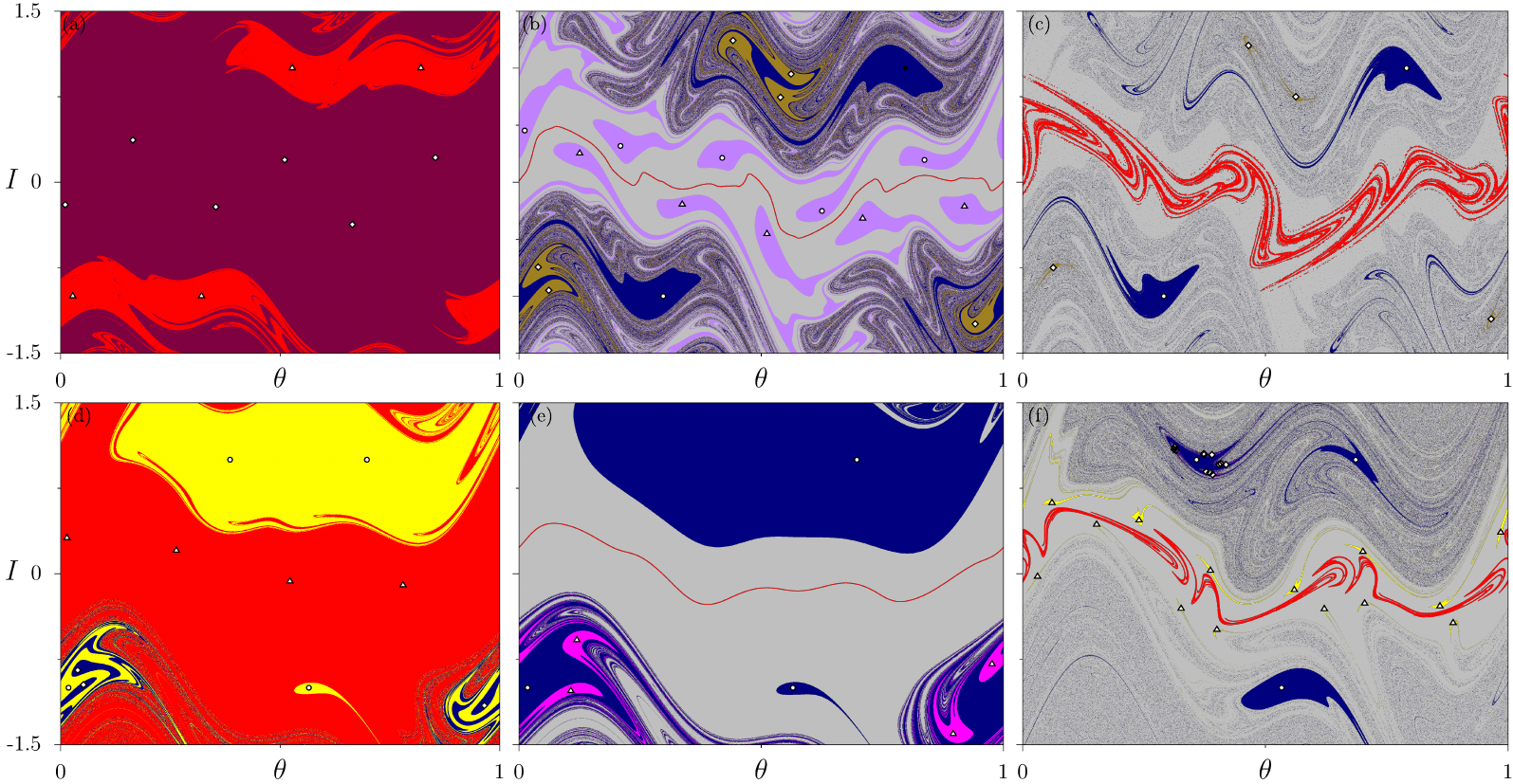}
     \caption{Basins of attraction are depicted, where each color represents a distinct basin. The parameter $\gamma$ is constant at $0.1$. In the top row: (a) $a = 0.36$ and $b = 0.15$, (b) $a = 0.22$ and $b = 0.3$, and (c) $a = 0.222646$ and $b = 0.4$, all for $\eta = 3$. In the bottom row: (d) $a = 0.26$ and $b = 0.3$, (e) $a = 0.2165$ and $b = 0.3$, and (f) $a = 0.32644$ and $b = 0.5$, all for $\eta = 2$.}
     \label{fig:9}
\end{figure*}
\section{Shearless attractors scenarios}\label{section5}

In the previous section, we analyzed the shearless attractor stability using the Lyapunov exponent and the bifurcation diagram perspective. This analysis was conducted for a fixed value of $b$ and a narrow range of $a$. With $\gamma$ set to $0.1$, the behavior of the DLM is influenced by both $a$ and $b$. When exploring bifurcation routes in systems with two parameters, intricate paths within the parameter space can be uncovered. In this section, our goal is to investigate the characteristics of solutions within the phase space for every parameter pair $(a,b)$. While bifurcation diagrams are commonly used for studying bifurcation routes when a single parameter is varied, extending this analysis to systems with two varying parameters presents challenges in visualization and interpretation. Therefore, we will focus on the Lyapunov exponent and the orbit period. We will calculate the Lyapunov exponent and the orbit period for each pair $(a,b)$, where $a$ ranges from $0$ to $1$ and $b$ ranges from $0$ to $0.6$. The results will be depicted using a color scale in the parameter space $a \times b$.

We use a systematic approach to create the parameter spaces shown in Figure \ref{fig:8} for $\eta=2$ and $\eta=3$ (a) and (b) respectively. Initially, we set a fixed value for the parameter $a$ and designate the indicator point of the conservative case, denoted as $(\textbf{\textrm{IP}}_1)_{1}$, as our IC for the first value of $b$. We repeat this process until the final iteration $n=2\times10^{5}$, excluding the initial $1\times10^{5}$ iterations. During this iteration, we calculate the orbit period up to a maximum period of $100$ with a precision of $1\times 10^{-8}$. For cases that did not correspond to any period, we numerically determined the largest Lyapunov exponent $\lambda$ and classified the system's behavior at that point. It is worth recalling that for maps, the quasi-periodic stability is characterized by the largest Lyapunov exponent remaining equal to zero while parameters vary. For the next value of $b$, we use the last iteration $(\theta, I)$ from the previous parameter as the initial condition, repeating the computation of $\gamma$ until $b=0.6$. This process is repeated for subsequent values of $a$, where the initial condition is reset. The parameter spaces shown in Figure \ref{fig:8} are generated using a linear grid comprising $2000\times2000$ points $(a,b)$. 

We generated Fig.~\ref{fig:8} intending to study the periodic structures in terms of period formation rules, highlighting the multi-stability character of the map dynamics, and comparing with the parameter space of the shearless breakup of Fig.~\ref{fig:3}. In both Fig.~\ref{fig:8}\textcolor{blue}{a}  and Fig.~\ref{fig:8}\textcolor{blue}{b}, corresponding to $\eta=2$ and $\eta=3$, respectively, the quasi-periodic regions (in black) bear a resemblance to the red regions in Figure 3. This shows that in scenarios where the shearless curve exists with the introduction of dissipation, it transforms into a quasi-periodic attractor. Conversely, a shift to a chaotic attractor (gray) occurs in regions of the breakup space where the shearless curve is absent (white).

Upon closer examination of small values of $a$, a difference becomes apparent between the parameter spaces for the shearless curve and the shearless attractor. In the conservative scenario, the shearless curve is observable, whereas in the dissipative case, there is a tendency for orbits to converge towards periodic attractors. It is important to note that while there is a preference for a single attractor in the phase space, rigorous proof of its existence remains elusive. As $a$ increases within the parameter space, a noticeable inclination towards chaotic behavior in the phase space becomes evident. Red, orange, and purple dots have been selected to indicate regions of positive Lyapunov.

Still, in Fig.\,\ref{fig:8}, we see periodic structures with a shape similar to a tongue, known as Arnold's tongues \cite{Arnold1991, Belair1985, MacKay1987}, starting in the quasi-periodic region (black) and ending in the chaotic (gray). The period formation rule for these structures is according to a Farey tree. Another characteristic identified in the parameter space in Fig. \ref{fig:8} is the region of multi-stability, which represents the mix of periodic structures.

\section{Multi-stability and basin entropy}\label{section6}

The DLM exhibits multi-stability under specific parameter values. As illustrated in Fig. 10, the phase spaces demonstrate the presence of multiple attractors and their corresponding colored basins of attraction for $\eta = 3$ at the top and $\eta = 2$ at the bottom. It is worth noting that for odd values of $\eta$, the periodic attractors are symmetric and possess a twin attractor that conforms to the transformation $T(I, \theta) = (-I, \theta + 0.5)$. However, this behavior is not observed for even values of $\eta$, indicating a lack of symmetry in the map.

The method used to distinguish between different scenarios in Fig.\,\ref{fig:9} is basin entropy \cite{Daza2016}. This method measures the intrinsic uncertainty within the basins by analyzing the entropy in the discretized phase space for a particular set of parameter values, each associated with $N_A$ distinguishable attractors. A finite number of boxes are placed in the phase space to discretize it, resulting in a two-dimensional grid with $N_T$ non-overlapping boxes. Each box contains initial conditions that converge to one of the $N_A$ attractors. After constructing the basins of attraction, colored points are found inside the boxes. The Gibbs entropy can be calculated for each box $i$.

\begin{equation}\label{entropy}
    S_{i} = \sum^{n_i}_{j=1}p_{i,j}\log \left(\dfrac{1}{p_{i,j}}\right),
\end{equation}

In this scenario, $n_i$ indicates the number of distinct colors within the box, where $n_i$ ranges from 1 to $N_A$. Conversely, $p_{i,j}$ represents the probability of a particular color $j$ appearing in box $i$. This probability is calculated by dividing the count of points with color $j$ by the total number of initial conditions in the box. It is important to note that $S_i$ attains a non-zero value only when more than one color is present in the box. For the purposes of this investigation, we focus on a box containing $25$ initial conditions.

After computing the entropy \eqref{entropy} for all $N_T$ boxes, we calculate the basin entropy $S_b$ and the boundary basin entropy $S_{bb}$. These quantities are defined as

\begin{equation}
    S_b = \dfrac{S}{N_T}= \dfrac{1}{N_T}\sum^{N_T}_{i=1}S_i
\end{equation}

for the basin entropy and

\begin{equation}
    S_{bb} = \dfrac{S}{N_b}= \dfrac{1}{N_b}\sum^{N_T}_{i=1}S_i,
\end{equation}

for boundary basin entropy, denoted as $S_{bb}$, is calculated using the number of boxes $N_b$ in the boundary between the basins, which are essentially boxes containing more than one color. This entropy measure can be used to characterize structures in the phase space and is linked to the uncertainty of the basin. The boundary values for basin entropy are $S_b = 0$  for a single attractor and $S_b = \log N_A$ for a fully randomized basin. Additionally, the boundary basin entropy is associated with the uncertainty of the boundaries.

\begin{table}
    \centering
    \begin{tabular}{c c c c c c}
    \hline
    \hline
         & $S_b$ & $S_{bb}$ & $N_a$ & Types of attractor\\
         \hline
         Fig.~\ref{fig:9}\textcolor{blue}{a} &  $0.0612$ & $ 0.512$ & 4 & P\\
        Fig.~\ref{fig:9}\textcolor{blue}{b} &  $0.163$ & $ 0.552$ & 9 & P,\,QP\\ 
       Fig.~\ref{fig:9}\textcolor{blue}{c} &  $0.252$ & $0.267$ & 5 & P,\,C\\
       Fig.~\ref{fig:9}\textcolor{blue}{d} &  $0.107$ & $0.508$ & 6 & P\\
       Fig.~\ref{fig:9}\textcolor{blue}{e} &  $0.0729$ & $0.533$ & 5 & P,\,QP\\
       Fig.~\ref{fig:9}\textcolor{blue}{f} &  $0.321$ & $0.391$ & 7 & P,\,C\\
       \hline
       \hline
    \end{tabular}
    \caption{Basin entropy $(S_b)$, boundary basin entropy $(S_{bb})$, number of attractors $(NA)$, and attractor types for each case of an indicated in Fig.~\ref{fig:9}. The letters P, QP, and C in the last column mean periodic, shearless quasi-periodic, and shearless chaotic, respectively.}
    \label{tab1}
\end{table}

Figure~\ref{fig:9} illustrates the basin of attraction and its attractors within the multi-stability scenarios of the DLM. The upper part of the figure corresponds to $\eta=3$, while the lower part corresponds to $\eta=2$, with all depicted scenarios demonstrating multi-stability. Figures~\ref{fig:9}\textcolor{blue}{a}  and \ref{fig:9}\textcolor{blue}{d} show scenarios devoid of shearless attractors, featuring only periodic attractors. Figures~\ref{fig:9}\textcolor{blue}{b} and \ref{fig:9}\textcolor{blue}{e} display phase spaces for these parameters, presenting quasi-periodic shearless attractors (in red) and their respective basins of attraction (in gray), along with periodic attractors and their basins. Lastly, Figures~\ref{fig:9}\textcolor{blue}{c} and \ref{fig:9}\textcolor{blue}{f} depict scenarios featuring chaotic shearless attractors (in red) and their respective basins (in gray) alongside periodic attractors.

For all the phase spaces present in Fig.~\ref{fig:9}, we compute the basin entropy $S_b$ and the boundary basin entropy $S_{bb}$ for each case using a grid comprising $2000 \times 2000$ boxes. We conduct $25$ iterations of initial conditions within each box, discarding the first $1\times10^5$ iterations as transient, and continue until $2\times10^5$ iterations to determine the final state or attractor. Table 1 displays the $S_b$ and $S_{bb}$ values. Based on the findings presented in Table~\ref{tab1}, it is evident that all basins exhibit a level of uncertainty when $S_b > 0$ in all instances. Notably, in Fig.~\ref{fig:9}\textcolor{blue}{c} and \ref{fig:9}\textcolor{blue}{f}, the area of mixed basins is more extensive and is characterized by high values of $S_b$. On the other hand, the lowest basin entropy values are observed in Fig.~\ref{fig:9}\textcolor{blue}{a} and \ref{fig:9}\textcolor{blue}{e}, indicating a lower degree of mixing in these cases.

For all the phase spaces presented in Fig.~\ref{fig:9}, we compute the basin entropy \( S_b \) and the boundary basin entropy \( S_{bb} \) for each case using a grid comprising \( 2000 \times 2000 \) boxes. We conduct $25$ iterations of initial conditions within each box, discarding the first \( 1 \times 10^5 \) iterations as transient, and continue until \( 2 \times 10^5 \) iterations to determine the final state or attractor. Table~\ref{tab1} displays the \( S_b \) and \( S_{bb} \) values. Based on the findings presented in Table~\ref{tab1}, it is evident that all basins exhibit some level of uncertainty, as indicated by \( S_b > 0 \) in all instances. Notably, in Figs.~\ref{fig:9}\textcolor{blue}{c} and \ref{fig:9}\textcolor{blue}{f}, the area of mixed basins is more extensive and characterized by high values of \( S_b \). Conversely, the lowest basin entropy values are observed in Figs.~\ref{fig:9}\textcolor{blue}{a} and \ref{fig:9}\textcolor{blue}{e}, indicating a lower degree of mixing in these cases.

When examining the basin boundary entropy, it can be divided into two scenarios: when \( S_{bb} < 0.5 \) and when \( S_{bb} > 0.5 \). In the former scenario, as depicted in Figs.~\ref{fig:9}\textcolor{blue}{c} and \ref{fig:9}\textcolor{blue}{f}, a distinct solid-colored region around \( I = 0 \) can be observed at the center of the basin, while a blending of basins occurs for higher and lower values of \( I \). In the latter scenario, a solid central region is observed around the periodic attractors in Figs.~\ref{fig:9}\textcolor{blue}{a}and \ref{fig:9}\textcolor{blue}{d}, as well as the shearless attractor in Figs.~\ref{fig:9}\textcolor{blue}{b} and \ref{fig:9}\textcolor{blue}{e}. The results for the first scenario are attributed to the solid region around \( I = 0 \), which reduces the final value of \( S_{bb} \). This is also a consequence of the large mixing region, increasing the number of boxes and decreasing the value of \( S_{bb} \). In contrast, the second scenario exhibits a relatively small mixing area, resulting in a smaller number of boxes and a larger value of \( S_{bb} \).

\section{Arnold tongues and shrimp-shaped structures}\label{section7}

\begin{figure*}
     \centering
    \includegraphics[width=0.8\textwidth]{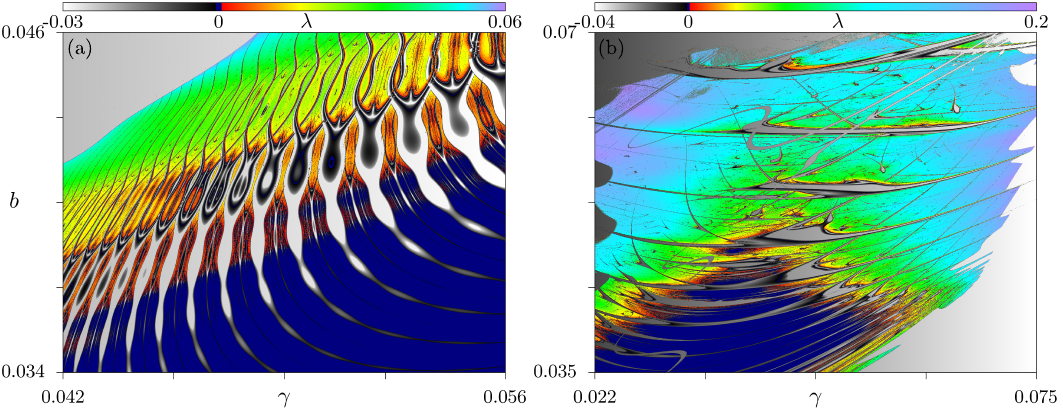}
     \caption{Parameter spaces $(\gamma, b)$ for (a) $\eta = 4$ and (b) $\eta = 7$. The colors represent the values of the largest Lyapunov exponent. The gradient from white to black indicates periodic orbits, navy blue indicates quasi-periodic orbits, and the gradient from red to purple indicates chaotic orbits. In both parameter spaces, periodic structures known as Arnold tongues are observed. These structures emerge from chaotic regions and extend into quasi-periodic regions.}
     \label{fig:10}
\end{figure*}

\begin{figure*}
     \centering
    \includegraphics[width=0.8\textwidth]{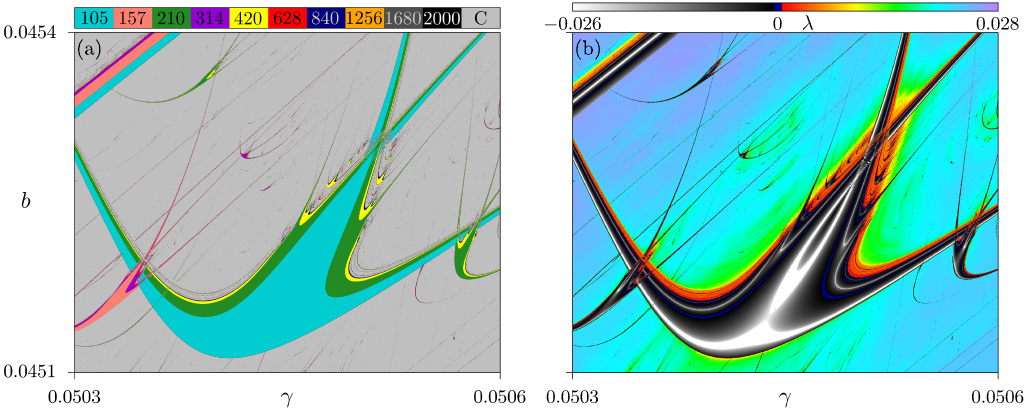}
     \caption{Magnifications of the parameter space $(\gamma, b)$ from Fig.~\ref{fig:10}\textcolor{blue}{a}. In (a), the colors represent the orbit period, counted up to $2000$. Chaotic orbits are denoted by (C) and shown according to the color legend. In (b), the colors represent the largest Lyapunov exponent. The gradient from white to black indicates periodic orbits, navy blue indicates quasi-periodic orbits, and the gradient from red to purple indicates chaotic orbits.}
     \label{fig:11}
\end{figure*}

\begin{figure*}
     \centering
    \includegraphics[width=0.8\textwidth]{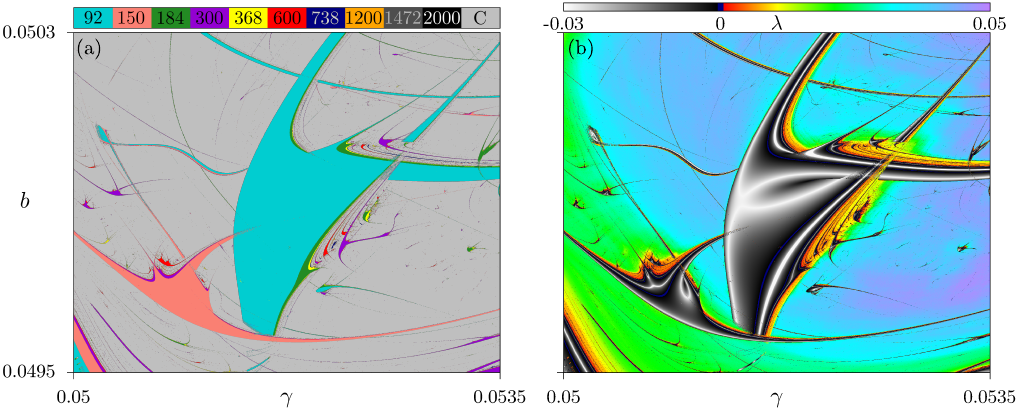}
     \caption{Magnifications of the parameter space $(\gamma, b)$ from Fig.~\ref{fig:10}\textcolor{blue}{b}. In (a), the colors represent the orbit period, counted up to $2000$. Chaotic orbits are denoted by (C) and shown according to the color legend. In (b), the colors represent the largest Lyapunov exponent. The gradient from white to black indicates periodic orbits, navy blue indicates quasi-periodic orbits, and the gradient from red to purple indicates chaotic orbits.}
     \label{fig:12}
\end{figure*}
In the previous sections, we focused on the behavior of the shearless attractors in terms of the parameters $\eta$, $a$, and $b$. For such analysis, we fixed the positions of the primary resonances $r_{1,2}$ and the dissipation $\gamma$. Now, our focus shifts to detecting periodic structures immersed in chaotic regions, specifically Arnold tongues and shrimp structures. This section presents the numerical results from iterating the DLM and qualitatively characterizing its dynamics. We graphically represent the parameter spaces $(\lambda, b)$ using the largest Lyapunov exponent and orbit period to achieve this.

Fig.\,\ref{fig:10} shows the parameter spaces $(\gamma, b)$, where the color scheme indicates the values of the largest Lyapunov exponent depicted in color for (a) $\eta = 4$ and (b) $\eta = 7$. The Lyapunov exponent was computed up to the final iteration $n = 2 \times 10^{5}$, excluding the initial $1 \times 10^{5}$ iterations as transient. The initial conditions used were $(\theta, I)=(0.25,0)$ for $\eta=4$; and $(\theta, I)=(\mathrm{IP}_1)_3$ for $\eta=7$, given by Eq.\eqref{IP}. In both parameter spaces, periodic, quasi-periodic, and chaotic scenarios are identified. quasi-periodic scenarios are shown in black, chaotic scenarios range from red to purple, and periodic scenarios are represented in a spectrum from white to gray, indicating a negative Lyapunov exponent. In both parameter spaces, periodic structures, known as Arnold tongues, are formed. Figs.~\ref{fig:11} and \ref{fig:12} present magnifications of Fig.~\ref{fig:10} for $\eta=4$ and $\eta=7$, respectively. Figs.~\ref{fig:11}\textcolor{blue}{a} and \ref{fig:12}\textcolor{blue}{a} depicts the orbit period using a color scheme, while Figs.~\ref{fig:11}\textcolor{blue}{b} and \ref{fig:12}\textcolor{blue}{b} illustrate the largest Lyapunov exponent, also employing a color scheme, for $\eta=4$ and $\eta=7$, respectively. In these magnifications, we observe primary shrimp-like structures that appear alongside additional shrimps in the vicinity of the main structures. In Fig.~\ref{fig:12}, we observe that the shrimp with a period of 150 overlaps with the main shrimp, which is an indicator of multi-stability in the region.

In the periodicity region, the two white curves depicted in Fig.~\ref{fig:11}\textcolor{blue}{b} and Fig.~\ref{fig:12}\textcolor{blue}{b} within the shrimp structures represent super-stable orbits, periodic orbits characterized by very high stability. These curves indicate that trajectories rapidly converge to attractors. In Fig.~\ref{fig:11}\textcolor{blue}{a} and Fig.~\ref{fig:12}\textcolor{blue}{b}, the orbit period parameter space reveals that the boundary separating the light blue and gray (chaotic) regions is a result of a tangent bifurcation. Similarly, the boundary between the light blue and green regions, as well as the subsequent boundaries (green to yellow regions), indicate period-doubling bifurcations. We can observe that at the borders of these regions, when compared with the parameter space of the Lyapunov exponent, the exponent passes through zero, indicating a bifurcation. This series of bifurcations outlines the route to chaos through a cascade of period-doubling bifurcations. 

Another feature of the shrimp structures that can be observed in the orbit period parameter space is their self-similarity at the secondary shrimps close to the legs of the main shrimp. In Fig.~\ref{fig:11}\textcolor{blue}{a} and Fig.~\ref{fig:12}\textcolor{blue}{a} the primary shrimp structure exhibits secondary shrimp structures with periods that are multiples of the primary period. These secondary shrimp structures can also include the presence of chaos signatures with periods that are multiples of a smaller base period. Additionally, a period-adding sequence can be observed, where the period increases by the primary period value, maintaining the periodicity of the main shrimp. This self-similarity characteristic can be observed across different primary periods, showing consistent patterns in the secondary structures.

\section{Conclusions}\label{section8}

The introduction of dissipation in the Labyrinthic map produces the transition of the shearless curve into a shearless attractor, regardless of the map symmetry in terms of $\eta$. This formed attractor can manifest as either quasi-periodic or chaotic and for the previous, possess the same frequency as the torus in the conservative case. Our analysis of the parameter spaces $(a,b)$ for the Lyapunov exponent reveals that quasi-periodic shearless attractors exhibit a structure similar to the parameter space associated with the breakup of the shearless curve in conservative maps for both odd and even values of $\eta$. For lower values of $a$ and $b$, the dissipation causes the ICs to asymptotically converge to periodic attractors or the quasi-periodic shearless attractor. In contrast, for higher values of $a$ and $b$, the ICs tend to converge towards the chaotic shearless attractor. We also observe structures resembling Arnold's tongues within these parameter spaces, which appear to conform to the period rule of the Farey sequence.

Additionally,through bifurcation diagrams and phase space constructions, we identify a transition to chaos for the shearless attractor. By fixing the parameter $b$, we observe the occurrence of a transition known as the Curry-Yorke route, within a specific range of $a$, during which the quasi-periodic attractor on the torus transitions into chaotic bands before finally evolving into a chaotic attractor.

Our investigation also explores the multi-stability within the system, which is observed in both symmetric and non-symmetric cases. By constructing the basins of attraction, we present the coexistence of three distinct attractors in the phase space: the shearless attractor and two twin periodic attractors, notably, the twin attractors arise specifically in the symmetric case. We analyze basin entropy and boundary basin entropy by discretizing the phase space for a set of parameter values, revealing that the basins exhibit a degree of uncertainty. Moreover, phase spaces with the same number of distinct attractors can present different interaction scenarios between the basins of attraction, which are discernible through basin entropy and boundary basin entropy analyses.

Finally, our examination of the parameter space $(\gamma, b)$ reveals common periodic structures characteristic of dissipative systems, including Arnold's tongues and shrimps. Through period counting and Lyapunov exponent analysis, we identify two bifurcation pathways leading to chaos within the shrimp structures. Moreover, our study uncovers super-stability curves and self-similarities within these formations, concluding that such structures can emerge regardless of their symmetry.

\section*{Acknowledgments}

The authors thank the financial support from the Brazilian Federal Agencies (CNPq) under Grant Nos. 304616/2021-4, 403120/2021-7, 301019/2019-3; the São Paulo Research Foundation (FAPESP, Brazil) under Grant No. 2018/01321-6, and 2019/07329–4; and support from Coordenação de Aperfeiçoamento de Pessoal de Nível Superior (CAPES) under Grants No. 88887.843528/2023-00. We extend our gratitude to Professor Jason Alfredo Carlson Gallas for his contributions to the field of nonlinear dynamics.

\section*{data availability}
The data that support the findings of this study are available from the corresponding author upon reasonable request.

\section*{References}
\bibliography{aipsamp}

\end{document}